\documentclass[11pt,notitlepage]{article}
\usepackage[a4paper, left=2cm, right=2cm, top=2cm, bottom=2cm]{geometry}
\usepackage[numbers]{natbib}
\usepackage{amsmath, amssymb, amstext}
\usepackage{graphicx}
\graphicspath{ {Plots/} }
\usepackage{ucs}
\usepackage[utf8x]{inputenc}
\usepackage[T1]{fontenc}
\usepackage[DIV=15]{typearea}
\usepackage{subcaption}
\usepackage{lettrine}

\begin{document}


\title{Observation of different reactivities of \emph{para-} and \emph{ortho-}water towards cold diazenylium ions}
\author{Ardita Kilaj$^{1\ast}$, Hong Gao$^{1,2\ast}$, Daniel R\"osch$^{1}$, Uxia Rivero$^{1}$, Jochen K\"upper$^{3,4,5,6}$ \\ and  Stefan Willitsch$^1$}
\date{}
\maketitle

\begin{center}
$^1$ Department of Chemistry, University of Basel, Klingelbergstrasse 80, 4056 Basel, Switzerland \\
$^2$ Present address: Beijing National Laboratory of Molecular Sciences, State Key Laboratory of Molecular Reaction Dynamics, Institute of Chemistry, Chinese Academy of Sciences, Beijing 100190, China \\
$^3$ Center for Free-Electron Laser Science, Deutsches Elektronen-Synchrotron DESY, Notkestrasse 85, 22607 Hamburg, Germany \\
$^4$ Department of Physics, Universit\"at Hamburg, Luruper Chaussee 149, 22761 Hamburg, Germany \\
$^5$ Department of Chemistry, Universit\"at Hamburg, Martin-Luther-King-Platz 6, 20146 Hamburg, Germany \\
$^6$ The Hamburg Center for Ultrafast Imaging, Universit\"at Hamburg, Luruper Chaussee 149,
22761 Hamburg, Germany \\
$\ast$ These authors contributed equally to the present work. \\
\end{center}


{\bf 
Water, H$_2$O, is one of the fundamental molecules in chemistry, biology and astrophysics. It exists as two distinct nuclear-spin isomers, \emph{para-} and \emph{ortho-}water, which do not interconvert in isolated molecules. The experimental challenges in preparing pure samples of the two isomers have thus far precluded a characterization of their individual chemical behaviour. Capitalizing on recent advances in the electrostatic deflection of polar molecules, we separated the ground states of \emph{para-} and \emph{ortho-}water in a molecular beam to show that the two isomers exhibit different reactivities in a prototypical reaction with cold diazenylium ions (N$_2$H$^+$). Based on \emph{ab initio} calculations and a modelling of the reaction kinetics using rotationally adiabatic capture theory, we rationalize this finding in terms of different rotational averaging of ion-dipole interactions during the reaction. The present results highlight the subtle interplay between nuclear-spin and rotational symmetry and its ramifications on chemical reactivity.
}
\newline{}



Water, H$_2$O, is one of the key molecules in nature, it acts as the fundamental solvent in biological systems and is one of the major molecular constituents of the universe. It exists in two forms, \emph{para(p)-} and \emph{ortho(o)-}water, which are distinguished by their values of the quantum number of the total nuclear spin $I$, where $I=0$ and 1 for \emph{p-} and \emph{o}-H$_2$O, respectively. Interconversion of the nuclear-spin isomers is forbidden in isolated molecules and nuclear-spin symmetry is usually conserved in collisions, by electromagnetic radiation and even in chemical reactions \cite{bunker98a, kanamori17a, quack77a}. However, nuclear-spin-symmetry interconversion has been observed in a variety of polyatomic molecules and has been rationalized to occur via doorway states with a mixed nuclear-spin character \cite{kanamori17a, sun05a, sun14a}. For water in the vapour and condensed phases, the \emph{para/ortho-}interconversion rates reported in the literature vary widely and remain a controversial topic \cite{tikhonov02a, mancatanner13a, georges17a}. 

Apart from their total nuclear spin, \emph{para-} and \emph{ortho-}water also differ in other important respects. Because the generalized Pauli principle dictates that the total molecular wavefunction has to be antisymmetric under the permutation of the two hydrogen nuclei in the molecule \cite{bunker98a}, \emph{(ortho-) para}-water is associated with (anti)symmetric rotational functions in the electronic-vibrational ground state of the molecule. Thus, nuclear-spin and rotational symmetry are intimately linked \cite{horke14a}. As a consequence, the ground state of \emph{para-}water is the absolute rotational ground state $|j_{K_aK_c}\rangle=|0_{00}\rangle$, whereas the ground state of \emph{ortho}-water is the first excited rotational state $|j_{K_aK_c}\rangle=|1_{01}\rangle$. Here, $j$ denotes the quantum number of the rotational angular momentum and $K_a$ and $K_c$ are the quantum numbers of the projection of the rotational angular momentum on the $a$ and $c$ principal axes of inertia of the molecule, respectively. 

Considering the different properties of its two nuclear-spin isomers and the eminent importance of water in a variety of chemical contexts, it begs the question whether \emph{para-} and \emph{ortho-}water also show different chemical behaviour. In a wider context, this problem ties into ongoing efforts to understand how different molecular degrees of freedom (translation, nuclear spin, rotation, vibration, electronic motion) and the interplay between them influence chemical reactivity. Despite the significant amount of studies focusing on vibrational effects in chemical reactions \cite{guettler94a, crim08a, liu16a}, the roles of nuclear spin and molecular rotation have scarcely been explored experimentally. This is mainly due to the fact that rotational energy transfer is likely to happen in any collision rendering it difficult to prepare molecules in specific rotational levels \cite{smith06a, chang15a} so that only comparatively few rotational-state resolved studies have been reported so far \cite{hauser15a, shagam15a, perreault17a}. A similar scarcity of data exists with regard to studies involving individual nuclear-spin isomers \cite{gerlich13a} which are in general difficult to separate and to prepare individually \cite{horke14a, kravchuk11a}.

Recent progress in manipulating polar molecules using electrostatic fields has made it possible to select and spatially separate different conformers and rotational states of molecules in supersonic molecular beams \cite{chang15a}. By combining this technology with a stationary reaction target of Coulomb-crystallized ions in a linear quadrupole ion trap (LQT) \cite{willitsch12a, willitsch17a}, we have recently studied conformer-selected molecule-ion reaction dynamics and observed that reaction-rate constants can strongly depend on molecular conformation \cite{chang13a, roesch14a}. Here, we extend this method to the separation of different nuclear-spin isomers for studies of ion-molecule reactions with control over the rotational and nuclear-spin state of the neutral reaction partner. As an example, we investigate the proton-transfer reaction of water with ionic diazenylium ($\rm N_2H^+$),
\begin{equation}
\rm H_2O + N_2 H^{+} \rightarrow N_2 + H_3O^{+},
\label{eq:reaction}
\end{equation}
an important molecule in astrochemistry which has been observed in the interstellar medium \cite{turner74a}. Its detection has proven crucial to trace molecular nitrogen in pre-stellar clouds to understand the early stages of star formation \cite{caselli02a, bergin02a}.

\begin{figure}[hb!]
	\centering
		\includegraphics[width=0.8\textwidth]{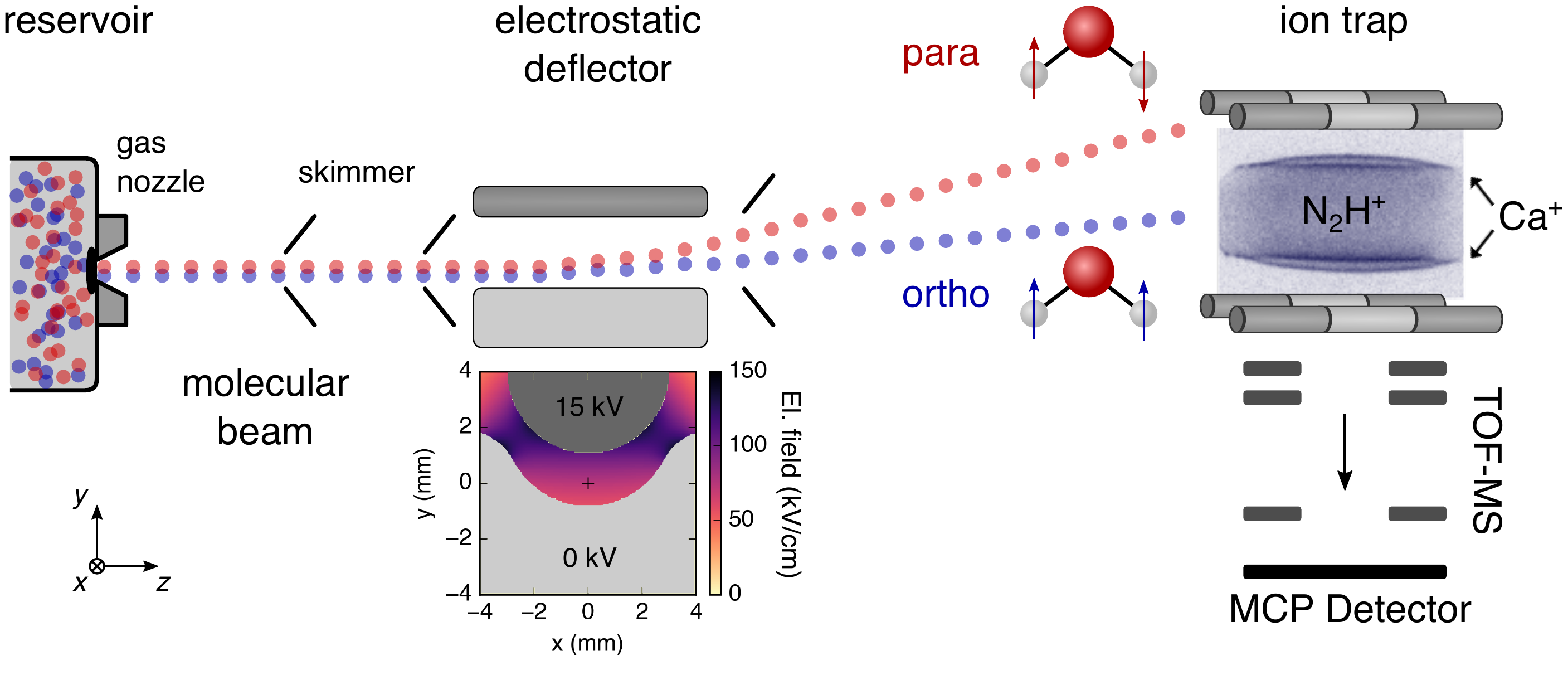}
		\caption{\textbf{Schematic of the experimental setup.} A pulsed molecular beam of water molecules seeded in argon emanates from a room-temperature reservoir through a pulsed gas nozzle and passes an electrostatic deflector. The inhomogeneous electric field inside the deflector (shown in the inset below) spatially separates \emph{para-} and \emph{ortho-}water molecules due to their different effective dipole moments. After the deflector, the beam is directed at an ion trap containing a Coulomb crystal of $\rm Ca^+$ and sympathetically cooled $\rm N_2H^+$ reactant ions (inset image). The products and kinetics of reactive collisions between $\rm N_2H^{+}$ and H$_2$O are probed using a time-of-flight mass spectrometer (TOF-MS) \cite{roesch16a}.} 
\label{fig:expsetup}	
\end{figure}


\section*{Results}
\label{sec: results}

The two ground states of \emph{para-} and \emph{ortho-}water show distinct responses to an electric field, i.e., different Stark-energy shifts and correspondingly different effective space-fixed dipole moments (Supplementary section "Monte Carlo Trajectory Simulations" and \cite{horke14a}). This enables their spatial separation by the electric field gradient of an electrostatic deflector \cite{horke14a, chang15a}. 

The experimental setup is schematically depicted in Figure \ref{fig:expsetup}. It consists of a molecular-beam machine equipped with the electrostatic deflector connected to an ultrahigh-vacuum chamber housing an ion trap \cite{roesch14a, roesch16a}. A beam of internally cold molecules was formed in a pulsed supersonic expansion of water seeded in argon carrier gas (Methods). The molecular beam was collimated by two skimmers before entering the electrostatic deflector. A voltage of 15 kV was applied across the deflector electrodes in order to generate a vertical electric field gradient for the spatial separation of the two nuclear-spin isomers. After passing another skimmer, the beam was directed towards a LQT. The trap was loaded with Coulomb crystals of laser-cooled $\rm Ca^{+}$ ions \cite{willitsch12a, chang13a} as well as sympathetically cooled N$_2$H$^+$ reactant ions (image inset in Figure \ref{fig:expsetup}). By vertically tilting the molecular beam apparatus relative to the LQT, different regions of the deflected molecular beam were overlapped with the Coulomb crystals \cite{roesch14a}. The tilting angle of the molecular beam apparatus defined a deflection coordinate $y$ for molecules arriving at the trap centre. After exposure to the molecular beam for a variable time period, the Coulomb crystals were ejected into a high-resolution time-of-flight mass spectrometer (TOF-MS) \cite{roesch16a} for the mass and quantitative analysis of their constituents.

\begin{figure}[hb!]
	\centering
		\includegraphics[width=0.9\textwidth]{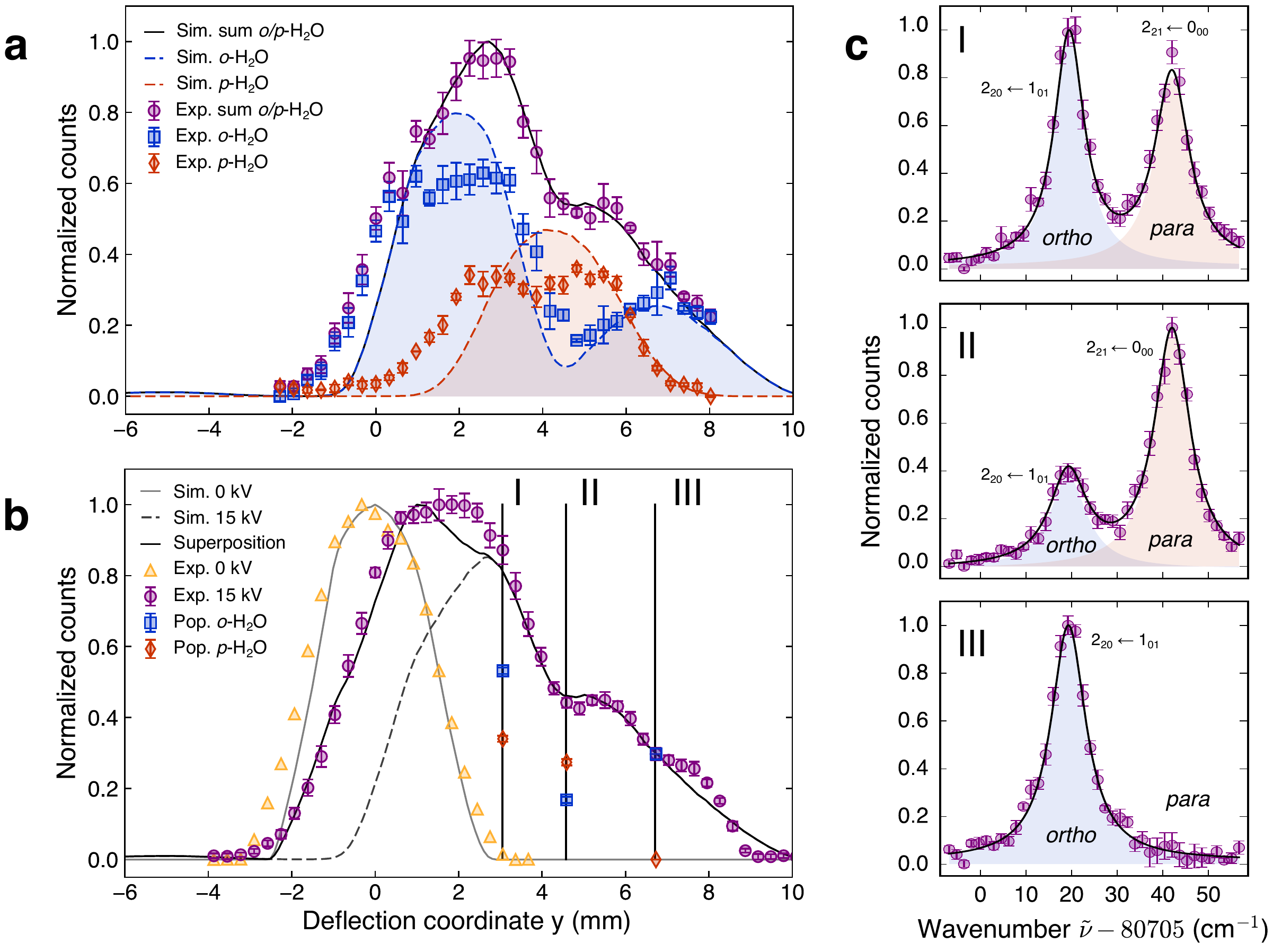}
 	\caption{\textbf{Comparison of the measured and simulated molecular-beam deflection profiles of the ground states of \emph{para-} and \emph{ortho-}water.} \textbf{a} Experimental isomer-specific density profiles of \emph{o-} (blue squares) and \emph{p-}H$_2$O (red diamonds) in the deflected molecular beam (deflector voltage 15 kV) measured by (2+1) REMPI together with the total deflection profile (sum of the \emph{ortho-} and \emph{para-}signals, purple circles). The lines represent Monte-Carlo trajectory simulations of the deflection profiles. The contributions from \emph{ortho-} and \emph{para-}water are indicated by the blue and red shaded areas, respectively. \textbf{b} Total water deflection profile measured by femtosecond-laser ionization for deflector voltages of 0 kV (yellow triangles) and 15 kV (purple circles). The three vertical lines marked I, II and III indicate the deflection coordinates at which reaction rates were measured. The red/blue symbols represent the relative populations of the isomers normalized to the total signal at positions I, II and III as determined from the REMPI spectra shown in c. \textbf{c} REMPI spectra of H$_2$O measured at the three positions I, II, III (purple circles). The two peaks observed at 80724 cm$^{-1}$ and 80747 cm$^{-1}$ correspond to transitions from the ground states of \emph{ortho-} and \emph{para-}water, respectively. The peaks are fitted with a sum of two Lorentzians (solid black line) with contributions from \emph{ortho-} and \emph{para-}isomers depicted as blue and red shaded areas, respectively. Error bars correspond to one standard error of at least three independent measurements.}
	\label{fig:deflection}
\end{figure}

In order to probe its composition in terms of quantum states and to characterize the spatial separation of the two nuclear-spin isomers, density profiles of the molecular beam were measured. A pulsed ultraviolet laser beam was used to generate water ions by $(2+1)$ resonance-enhanced multi-photon ionization (REMPI) via selected rotational levels of the $\tilde{\mathrm{C}}$ electronic state  \cite{yang10a}. The ions were subsequently ejected into the TOF-MS. This technique enabled the selective detection of the ground states of either \emph{para-} or \emph{ortho-}water and the determination of the individual density profiles of the two isomers in the beam. From a REMPI spectrum of an undeflected water beam, it was confirmed that the supersonic expansion was composed predominantly of the $j_{K_aK_c}=0_{00}$ and $1_{01}$ rotational states of H$_2$O, i.e., the ground states of \emph{p-} and \emph{o-}H$_2$O, respectively. A possible minor contribution from the $1_{10}$ state did not interfere with the present experiments (Supplementary sections "Analysis of REMPI spectra" and "Composition of the molecular beam").

Figure \ref{fig:deflection}a shows deflection profiles of \emph{para-} (red diamonds) and \emph{ortho-} (blue squares) water obtained from the ion signal at mass-to-charge ratio $m/z=18~u$ recorded as a function of the deflection coordinate $y$ at a deflector voltage of 15 kV. The purple circles represent the sum of the \emph{para-} and \emph{ortho-}profiles. The different projections of the angular momentum of the ground state of \emph{ortho}-water onto the space-fixed direction of the electric field leads to two components, $M=0$ and $|M|=1$, which exhibit a weak and strong Stark shift, respectively.
These correlate with the two peaks of the \emph{o-}H$_2$O deflection profile at low and high deflection coordinates, respectively. Contrarily, the \emph{para-}isomer only has one angular-momentum projection component $M=0$ with an intermediate Stark shift such that its deflection profile shows a single peak situated in between the two peaks of the \emph{ortho-}form. In this way, a partial spatial separation of the two isomers was achieved and the $o/p$-ratio was well defined at each deflection coordinate \cite{horke14a}. The solid and broken lines show corresponding theoretical deflection profiles derived from Monte Carlo trajectory simulations (Supplementary section "Monte Carlo trajectory simulations"). 

In addition, a femtosecond (fs) laser was employed to probe molecules reaching the trap centre by strong-field ionization irrespective of the species or the internal quantum state \cite{teschmitt17a}. Subsequent ion ejection into the TOF-MS enabled the determination of the combined relative density of \emph{para-} and \emph{ortho-}water molecules in the beam as a function of the deflection coordinate. The acquired beam profiles for deflector voltages of 0 and 15 kV are presented in Figure \ref{fig:deflection}b as yellow triangles and purple circles, respectively. At 15~kV, experiment and simulation (dashed grey line) agree well at large deflection coordinates, but differ significantly around $y=0$. In this region, the mass spectra indicate the presence of clusters formed in the supersonic expansion. A fs laser pulse can break these clusters resulting in water ions detected together with the water-monomer signal at $m/z=18~u$ in the TOF-MS. Our data also show that these clusters are not deflected and do not contaminate the beam at deflection coordinates larger than 2 mm (Supplementary section "Composition of the molecular beam"). This picture is corroborated by the reproduction of the experimental beam profile by a weighted superposition (black solid line) of simulations of the deflected water beam at 15 kV and an undeflected beam at 0 kV (grey solid line). 

Probing the specific reactivities of the two isomers requires the preparation of samples with well defined \emph{para/ortho-}ratios. Based on the deflection profiles and the simulations in Figure \ref{fig:deflection}a, three deflection coordinates with varying relative populations of \emph{para-} or \emph{ortho-}water were chosen. At each of these positions labelled I, II and III in Figure  \ref{fig:deflection}b, REMPI spectra were recorded from which the populations of \emph{para-} and \emph{ortho-}water were determined (Figure \ref{fig:deflection}c). From fits of the intensities of the lines in the spectra, the relative populations of the two isomers were obtained (Supplementary section "Analysis of REMPI spectra"). The populations of \emph{para-} (\emph{ortho-}) water thus obtained were 39(1)\% (61(1)\%), 62(2)\% (38(2)\%) and 0\% (100\%) at positions I, II and III. We note that at position III, the beam consists of pure \emph{ortho-}water within the measurement uncertainties and therefore enables a direct measurement of the reaction rate of \emph{ortho-}water.

\begin{figure}[hb!]
	\centering
		\includegraphics[width=0.6\textwidth]{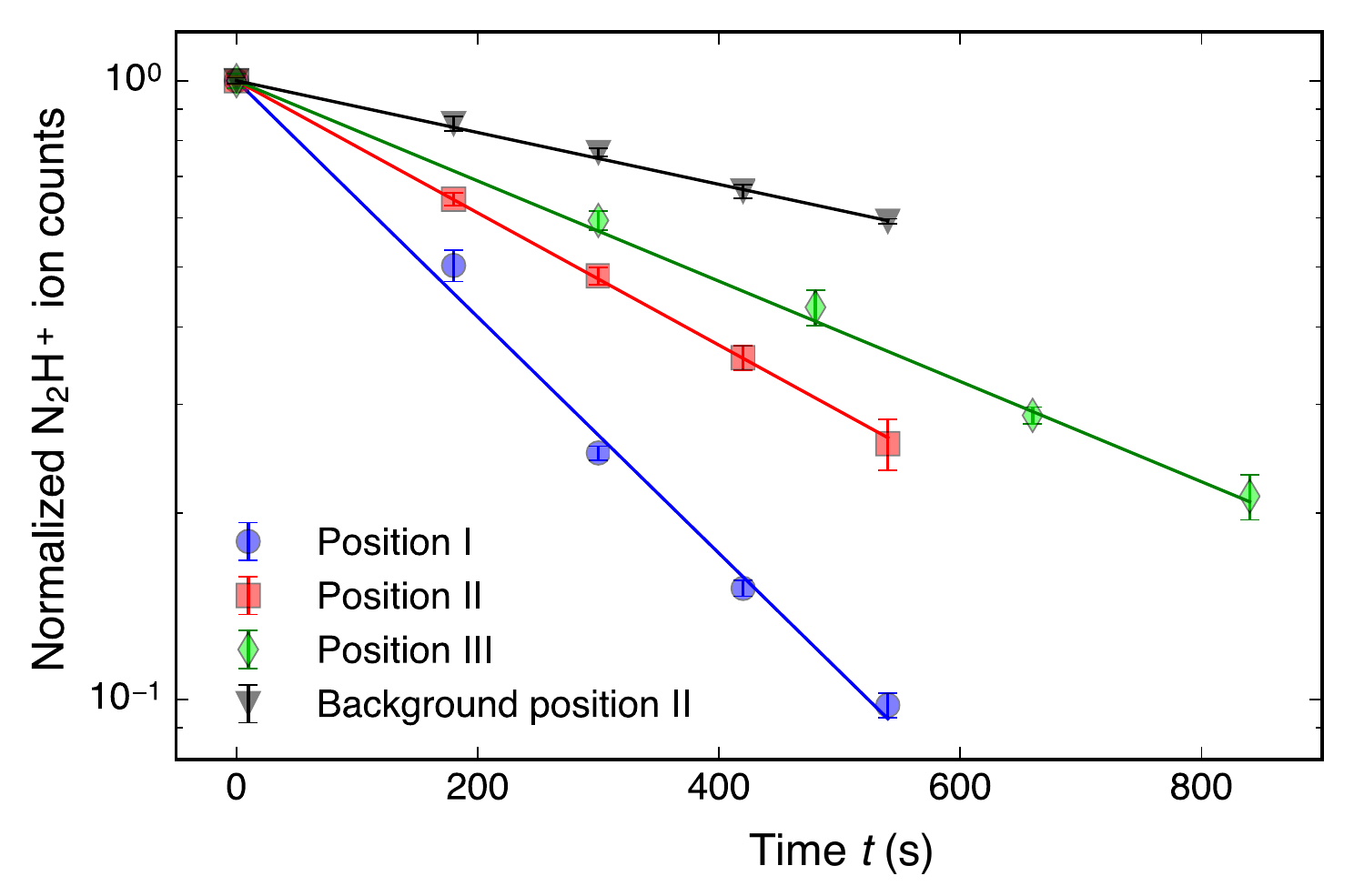} 	
		\caption{\textbf{Reaction-rate measurements at the deflection coordinates I--III indicated in Figure \ref{fig:deflection}b.} The data are normalized to the ion signal at time $t=0$. The lines represent fits to the data according to an integrated pseudo-first-order rate law. The black triangles show an example of a measurement of the reaction rate with background gas at position II for comparison. Error bars correspond to one standard error of four independent measurements.}
	\label{fig:rates}
\end{figure}

With the molecular beam prepared with well-known ratios of the two isomers at different deflection coordinates, measurements of the rate of reaction (\ref{eq:reaction}) were performed (Methods). First, a Coulomb crystal of about 1000 $\rm Ca^+$ ions was loaded into the LQT. Then, nitrogen gas was leaked into the vacuum chamber and N$_2^+$ ions were generated by fs-laser ionization. After the formation of $\rm N_{2}^{+}$ ions in the trap, H$_2$ gas was leaked into the vacuum chamber to quantitatively convert N$_2^+$ into N$_2$H$^+$ via the reaction $\mathrm{N_2^+ + H_2 \rightarrow N_2H^++H}$. The thus formed $\rm N_{2}H^{+}$ ions were sympathetically cooled into the Coulomb crystal and accumulated at its centre (inset in Figure \ref{fig:expsetup}). Subsequently, the molecular beam apparatus was set to a specific deflection coordinate and the deflector was turned on at a voltage of 15~kV to direct the molecular beam at the Coulomb crystal and engage the reaction. After a variable period of exposure to the molecular beam, the reduction of the number of N$_2$H$^+$ reactant ions was probed by ejecting the Coulomb crystal into the TOF-MS \cite{roesch16a} (Figure \ref{fig:rates}). In addition, the formation of $\rm H_3O^+$ as the ionic reaction product was verified using the TOF-MS. The rate measurements were repeated five times for each of the three deflection coordinates I--III. For every reaction measurement, a subsequent measurement of the rate of reaction of $\rm N_{2}H^{+}$ with the background gas in the vacuum chamber was performed by pointing the molecular beam away from the centre of the LQT at a deflector voltage of 0~kV. Since the number of water molecules is continuously replenished through the molecular beam, the rate constants could be determined within the framework of a pseudo-first-order kinetics treatment (Supplementary section "Reaction-rate constants"). The pseudo-first-order rate constants of the reactions with background gas were directly subtracted from the total rate constants to give the rate constants for the reactions of the diazenylium ions with water from the molecular beam.

Combining the determinations of the total reaction-rate constants, of the relative populations of \emph{o-} and \emph{p-}H$_2$O at positions I, II and III, and of the combined density profile of the deflected beam, the individual reaction-rate constants $k_o$ and $k_p$ of \emph{ortho-} and \emph{para-}water, respectively, were deduced. For each of the three deflection coordinates $y_i$ ($i \in \{\rm I, II, III\}$), the total first-order rate constant $k'_{\mathrm{tot},i}$ is given by
\begin{equation}
k'_{\mathrm{tot},i} = \tilde{n}_i \left(p_{o,i}\; k'_o + p_{p,i} \; k'_p\right),
\label{eq:rate_eq_system}
\end{equation}
where $p_{o,i}$ and $p_{p,i}$ are the populations of \emph{ortho-} and \emph{para-}water, respectively, and $\tilde{n}_i$ are the relative densities of the water beam at positions $y_i$. $\tilde{n}_i$ is obtained from the beam profile $I(y)$ measured with the fs laser at 15 kV (Figure \ref{fig:deflection}b) via $\tilde{n}_i = I(y_i)/I(y_\mathrm{II})$ with position II taken as reference point. Once $k'_o$ and $k'_p$ were known, the relative difference of the reaction rates of the two isomers was calculated as $r = 2(k'_p - k'_o)/(k'_p + k'_o)$.

The system of equations (\ref{eq:rate_eq_system}) was solved by a least-squares optimization. From the experimental data, the pseudo-first-order rate constants were determined to be $k'_o = 1.4(1)\times 10^{-3}~\mathrm{s^{-1}}$ and $k'_p = 1.8(1)\times 10^{-3}~\mathrm{s^{-1}}$ yielding a relative difference $r=23(9)\%$ between the reactivities of the \emph{para-} and \emph{ortho-}isomers. 

Absolute bimolecular reaction-rate constants $k_{o/p}$ were calculated from $k_{o/p}=k'_{o/p}/n_{\mathrm{H_2O, II}}$, where the total time-averaged beam density at the reference position II, $n_\mathrm{H_2O, II} = 3.0(5)\times 10^5~\mathrm{cm^{-3}}$, was estimated according to the procedures described in Supplementary section "Density of the molecular beam." Using this information, the absolute reaction rates obtained from this experiment are \linebreak $k_{o} = 4.8(9)\times 10^{-9}~\mathrm{cm^3s^{-1}}$ for \emph{ortho-}water and $k_{p} = 6(1) \times 10^{-9} ~\mathrm{cm^3s^{-1}}$ for \emph{para-}water, respectively.


\section*{Discussion}
\label{sec: discussion}

\begin{figure}[hb!]
	\centering
		\includegraphics[width=0.8\textwidth]{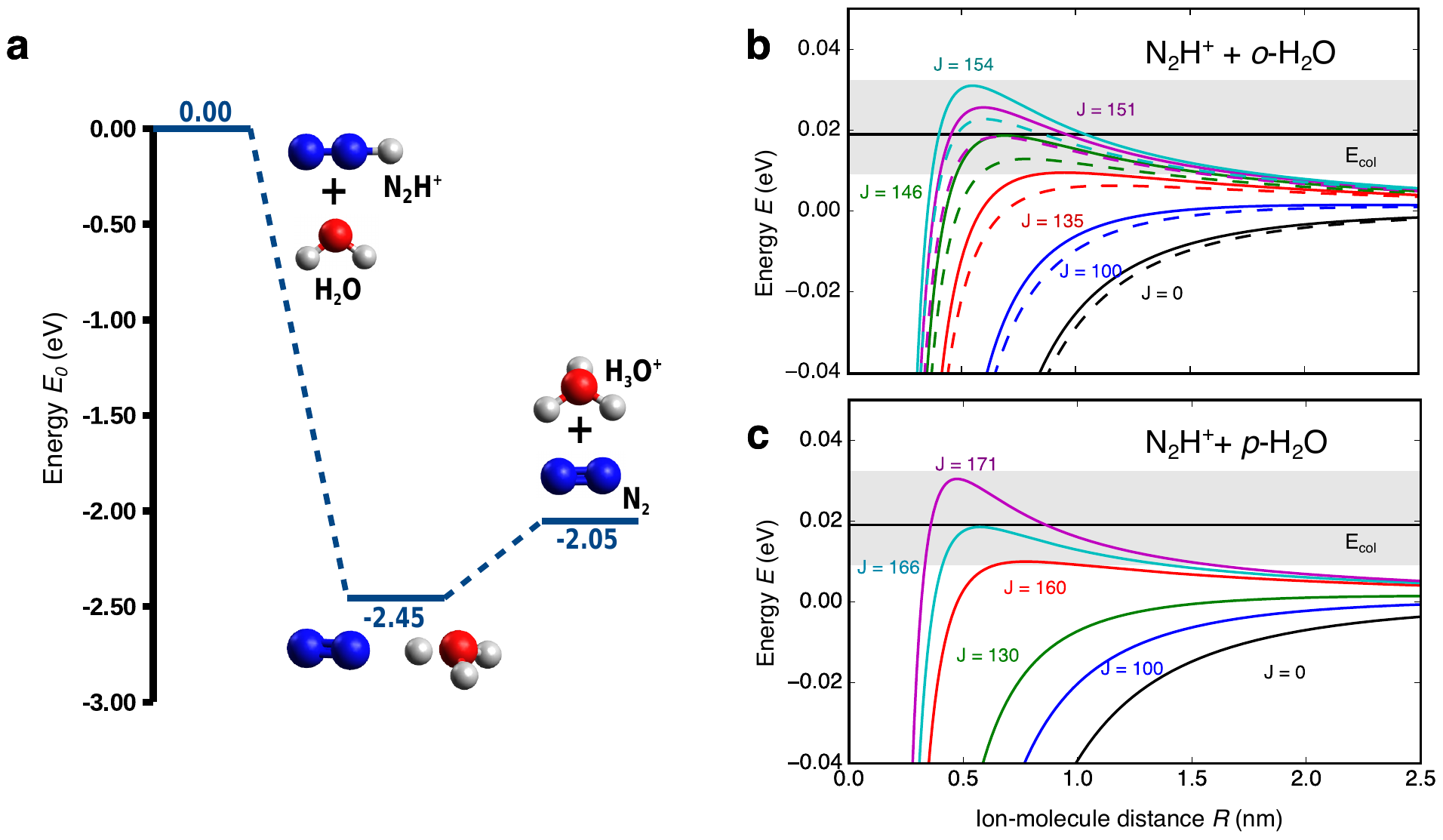}
		\caption{\textbf{Theoretical predictions from \emph{ab initio} calculations and adiabatic capture theory.} \textbf{a} Potential-energy profile along the reaction coordinate for the proton transfer reaction between N$_2$H$^+$ and H$_2$O at the CCSD/aug-cc-pVTZ level of theory. The relative energies with respect to the reactants as well as the structures of the stationary points are shown. Blue, red and white spheres represent nitrogen, oxygen and hydrogen atoms. 
\textbf{b,c} Rotationally adiabatic, centrifugally corrected long-range interaction potentials for the reaction of the ground states of \emph{o-} (b) and \emph{p-} (c) $\rm H_2O$ with $\rm N_2H^+$ for different values of the total angular momentum quantum number $J$. In b, the dashed (solid) lines correspond to the $|\Omega|=0 (1)$ components of the \emph{ortho-}ground state. The grey-shaded areas show an estimate of the uncertainty in the experimental collision energy $E_\text{col}$ indicated by the black horizontal line.} 
\label{fig:capture}	
\end{figure}

To understand the reason for the different reactivities of \emph{para-} and \emph{ortho-}water in the present case, \emph{ab initio} calculations of the energy profile of the reaction were performed (Methods and Supplementary section "Theory"). As can be seen in Figure \ref{fig:capture}a, the reaction was found to be barrierless and to proceed via the formation of an intermediate complex in which one hydrogen atom is shared between the nitrogen and water moieties. No transition state could be located within O-H$^+$ distances ranging from 100 to 260~pm. The energy of the products was found to be about 2~eV lower than the one of the  reactants. This situation suggests that the kinetics can be modelled within the framework of a rotationally adiabatic quantum capture theory for barrierless ion-molecule reactions \cite{clary87a, stoecklin92a}. According to this approach, the reaction rates are entirely dominated by the properties of the long-range interaction potential of the reactants and centrifugal effects. All collisions up to a maximum total angular momentum $J_\text{max}$ for which the relative kinetic energy exceeds the height of the centrifugal barrier lead to a successful reactive encounter (Supplementary section "Theory").

For the present case, the relevant terms in the long-range interaction potential are the charge-induced dipole and charge-permanent dipole interactions. In the current experiments, both nuclear-spin isomers were cooled down to their relevant rotational ground states. The anisotropic nature of the ion-dipole interaction implies that it is sensitive to the rotational quantum state of the neutral molecule. Figures \ref{fig:capture}b and c show rotationally adiabatic, centrifugally corrected interaction potentials for collisions of $\rm N_2H^+$ with \emph{o-} and \emph{p-}H$_2$O, respectively, as a function of the total angular momentum quantum number $J$. For the case of vanishing total angular momentum ($J=0$), one can see that the potential energy curves for the \emph{ortho-}species are less steep than the one of the \emph{para-}isomer, reflecting the stronger rotational averaging of the ion-dipole interaction in the ground state of $o$-H$_2$O. For the $|j_{K_aK_c}|\Omega|\rangle=|0_{00}0\rangle$ ground state of the \emph{para-}species one obtains a maximum collisional angular-momentum $J_\mathrm{max} = 166$ at the experimental collision energy $E_\text{col}=0.019$~eV. Here, $\Omega$ is the quantum number of the projection of $\Vec{j}$ on the distance vector between the ion and the neutral molecule in a body-fixed coordinate system describing the collision \cite{clary87a, stoecklin92a}. This value can be compared to $J_\mathrm{max} = 151$ and $J_\mathrm{max} = 146$ for the $|1_{00}0\rangle$ and $|1_{01}1\rangle$ states  of the \emph{ortho-}isomer, respectively. 

The quantum capture model predicts a reaction-rate constant of $k_p^\mathrm{AC} = 5(1)\times 10^{-9}~\mathrm{cm^3s^{-1}}$ for \emph{para}-water compared to $k_o^\mathrm{AC} = 4.0(9)\times 10^{-9}~\mathrm{cm^3s^{-1}}$ for \emph{ortho-}water. The value for the \emph{ortho-}species was obtained by summing over the contributions of all $\Omega$ components to the reaction cross section (Supplementary section "Theory"). Assuming that the original preparation of $M$ states of $o-$H$_2$O in the deflector was scrambled during the transit of the molecules through the RF fields in the ion trap. In a classical interpretation of this result, the higher maximum angular momentum obtained for the \emph{para-}isomer implies a larger impact parameter and, therefore, a higher reaction rate. The theoretical predictions agree with the measured values within their experimental uncertainties. The relative difference of the theoretical reaction-rate constants was calculated to be $r=24(5)\%$ which can be compared with the experimental value of $r=23(9)\%$. The errors in the calculated rate constants arise from the uncertainty in the collision energy due to the experimental velocity distribution of the molecular beam and the micromotion of the ions in the large Coulomb crystals \cite{willitsch12a} (Supplementary section "Collision velocity").

In conclusion, we have studied chemical reactions of the spatially separated ground states of \emph{para-} and \emph{ortho-}water with cold diazenylium ions. We found a 23(9)\% higher reactivity for the \emph{para} nuclear-spin isomer which we attribute to the smaller degree of rotational averaging of the ion-dipole long-range interaction compared to the \emph{ortho-}species. The observed difference in reactivities is thus a rotational effect which is induced by the nuclear-spin symmetry via the generalized Pauli principle. The present results highlight the subtle interplay between nuclear-spin and rotational symmetry and its ramifications on chemical reactivity. They also provide an illustration of the effects of exchange symmetry on chemical processes which may be put in context with, e.g., its manifestations in the dynamics of ultracold collisions \cite{ospelkaus10b}. Measurements such as the ones presented here fill a gap in experimental investigations of ion-neutral reactions with control over the quantum states of the reactants. Indeed, to our knowledge the present study is the first in which rotationally state-selected polyatomic neutral molecules have been reacted with ions. The methods employed here are applicable to studies of a broad range of ion-neutral process. They enable a quantitative understanding of how different molecular quantum states and, as demonstrated in our earlier study \cite{chang13a}, molecular conformations, influence chemical reactivity.


\section*{Methods}
\label{sec:methods}

(2+1)-resonance-enhanced multiphoton-ionization (REMPI) spectra of H$_2$O were acquired using the output of a frequency-doubled dye laser pumped by the 3rd harmonic (355 nm) of a Nd:YAG laser. The resulting laser pulses at 248 nm with an energy of about 1.5 mJ were focused into the molecular beam using a lens with a focal length of 30 cm. Details on the procedure for the analysis of the REMPI spectra are given in the Supplementary section "Analysis of REMPI spectra". 

Strong-field ionization of H$_2$O molecules was performed with pulses from a Ti:Sapphire femtosecond laser (CPA 2110, Clark-MXR, Inc.) at a wavelength of 775 nm and pulse duration of 150 fs focused down to a beam diameter of $\approx50~\mu$m. Similarly, Ca atoms as well as N$_2$ molecules  (partial pressure $4.0\times 10^{-9}$ mbar) were ionized using the fs laser before loading into the ion trap. The N$_2^+$ ions were reacted with H$_2$ gas leaked into the ultrahigh-vacuum chamber for 30~s at  a partial pressure of $p=2\times10^{-9}$~mbar to yield N$_2$H$^+$ ions. 

The ions were trapped in a radio frequency (RF) linear-quadrupole ion trap (LQT) operated at a peak-to-peak RF voltage $V_\mathrm{RF,pp} = 800$~V and frequency $\Omega_\mathrm{RF} = 2\pi \times 3.304~\mathrm{MHz}$. Doppler laser cooling of Ca$^+$ was achieved using two laser beams at 397 nm and 866 nm generated by frequency-stabilized external-cavity diode lasers \cite{willitsch12a}. Laser and sympathetic cooling as well as Coulomb crystallization of the ions were monitored by imaging the laser-cooling fluorescence of the Ca$^+$ ions onto a camera coupled to a microscope (see sample image inset in Figure \ref{fig:expsetup}). 

The LQT was connected to a TOF-MS orthogonal to the molecular-beam propagation axis for the mass and quantitative analysis of reactant and product ions \cite{roesch16a}.

The molecular beam was generated from water vapour at room temperature and seeded in argon carrier gas at 3 bar. The gas mixture was pulsed through an Amsterdam cantilever piezo valve (ACPV2, $150~\mu$m nozzle) at a repetition rate of 200 Hz and a pulse width of $30~\mu$s. The velocity of the resulting molecular beam was measured to be 575(65) m/s. The electrostatic deflector consists of a pair of 15.4 cm long electrodes separated by 1.4 mm \cite{roesch14a}.

Effective dipole moments and Stark energy curves of individual rotational states of water were calculated for input into Monte-Carlo trajectory simulations using the CMIstark software package \cite{chang14a}. Details of the numerical procedures are reported in the Supplementary section "Monte Carlo trajectory simulations".

\emph{Ab initio} calculations of the potential energy surface of the title reaction were performed using  the Gaussian 09 suite of codes \cite{g09}. Geometry optimizations of stationary points were carried out at the CCSD/aug-cc-pVTZ level of theory.


\section*{Author contributions}
D.R. designed and built the apparatus with help from H.G. A.K. and H.G performed the experiments. A.K analysed the data, performed theoretical calculations and simulated spatial beam profiles.  D. R. wrote the computer code for Monte-Carlo trajectory simulations and capture-theory calculations. U.R. performed the quantum-chemical calculations. A.K., H.G., J.K. and S.W. wrote the manuscript. S.W. and J.K. conceived and supervised the project. All authors have read and approved the final manuscript.

\section*{Acknowledgements}
We thank Philipp Kn\"opfel, Grischa Martin and Georg Holderried for technical support. Dominique Ostermayer and Jolijn Onvlee are acknowledged for their assistance with the experiments and simulations, respectively. This work is supported by the Swiss National Science Foundation under grant nr. BSCGI0\_157874.

\section*{Competing financial interests}
The authors declare no competing financial interests.


\bibliography{Main-Oct17,new_refs}
\bibliographystyle{apsrev}

\end{document}


\title{Supplementary Information: \\Observation of different reactivities of \emph{para-} and \emph{ortho-}water towards cold diazenylium ions}
\author{Ardita Kilaj$^{1\ast}$, Hong Gao$^{1,2\ast}$, Daniel R\"osch$^{1}$, Uxia Rivero$^{1}$, Jochen K\"upper$^{3,4,5,6}$ \\ and  Stefan Willitsch$^1$}
\date{}
\maketitle

\begin{center}
$^1$ Department of Chemistry, University of Basel, Klingelbergstrasse 80, 4056 Basel, Switzerland \\
$^2$ Present address: Beijing National Laboratory of Molecular Sciences, State Key Laboratory of Molecular Reaction Dynamics, Institute of Chemistry, Chinese Academy of Sciences, Beijing 100190, China \\
$^3$ Center for Free-Electron Laser Science, Deutsches Elektronen-Synchrotron DESY, Notkestrasse 85, 22607 Hamburg, Germany \\
$^4$ Department of Physics, Universit\"at Hamburg, Luruper Chaussee 149, 22761 Hamburg, Germany \\
$^5$ Department of Chemistry, Universit\"at Hamburg, Martin-Luther-King-Platz 6, 20146 Hamburg, Germany \\
$^6$ The Hamburg Center for Ultrafast Imaging, Universit\"at Hamburg, Luruper Chaussee 149,
$\ast$ These authors contributed equally to the present work. \\
\end{center}


\section{Analysis of REMPI spectra}
\label{sec:rempi_spectra}

\begin{figure}[h]
	\centering
		\includegraphics[width=0.6\textwidth]{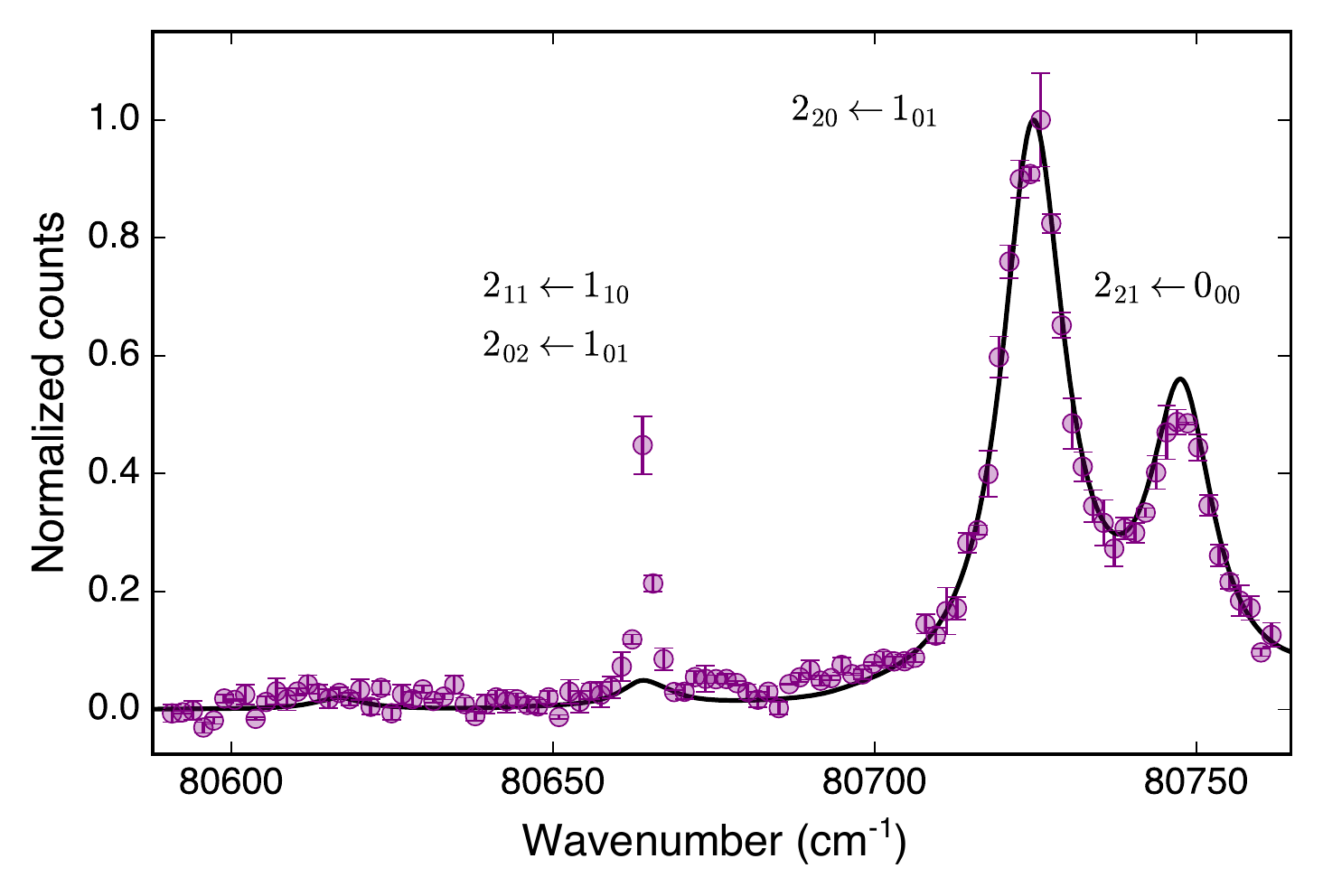}
 	\caption{(2+1) REMPI spectrum (data points) of a beam of water seeded in argon at a backing pressure of 3 bar and a  simulation of the spectrum (line) at a rotational temperature of 7 K. Uncertainties correspond to one standard error.}
\label{fig:SI_rempi_scan_0kV}
\end{figure}

In order to determine the rotational temperature of the undeflected water beam, a (2+1) REMPI spectrum was recorded by monitoring the ion signal detected at mass-to-charge ratio $m/z=18$~u in the TOF-MS as a function of the UV laser wavelength (Figure \ref{fig:SI_rempi_scan_0kV}). The experimental spectrum was compared to a simulation using the software package PGOPHER \cite{western16a}. Best agreement was found by assuming a rotational temperature of 7 K in line with previous studies \cite{yang10a, horke14a}. Two dominant transitions were identified, the $|j_{K_aK_c}\rangle=|2_{20}\rangle\leftarrow |1_{01}\rangle$ transition of \emph{ortho}-water at $80724~\mathrm{cm^{-1}}$ and the $|2_{21}\rangle\leftarrow |0_{00}\rangle$ transition of \emph{para}-water  at $80747~\mathrm{cm^{-1}}$. $j, K_a, K_c$ denote the rotational quantum numbers of an asymmetric top. The experimental intensities of the lines, i.e., the areas $A$ underneath the peaks, were determined from fits of a sum of Lorentz functions to the spectral features. 

The relative line strengths $S_\mathrm{o}$ and $S_\mathrm{p}$ of the two transitions for \emph{ortho-} and \emph{para-}water, respectively, were estimated from the relevant peak areas $A_\mathrm{o}$ and $A_\mathrm{p}$. Assuming that both nuclear-spin isomers were cooled down to their relevant ground states and that the \emph{ortho:para} ratio of 3:1 prevalent in the water vapour prior to expansion was conserved in the molecular beam, we set
\begin{equation}
3 = \frac{A_\mathrm{o}S_\mathrm{p}}{A_\mathrm{p}S_\mathrm{o}}.
\end{equation}
With $S_\mathrm{o} = 1$, we obtained $S_\mathrm{p} = 1.51(1)$. 

Finally, the populations $p_{\sigma}$  ($\sigma \in \{o, p\}$) at a given deflection coordinate $y_i$ ($i \in\{I,II,III\}$) were calculated from the peak areas $A_\sigma(y_i)$ and line strengths $S_\sigma$. From the scaled peak areas $\tilde{A}_\sigma(y_i) = A_\sigma(y_i) / S_\sigma$, we get
\begin{equation}
p_\sigma(y_i) = \frac{\tilde{A}_\sigma(y_i)}{\tilde{A}_\mathrm{p}(y_i) + \tilde{A}_\mathrm{o}(y_i)}.
\end{equation}
The populations obtained using this method are listed in Table \ref{tab:rates_equation_system}.

\section{Monte Carlo Trajectory Simulations}

\begin{figure}[h]
	\centering
		\includegraphics[width=0.8\textwidth]{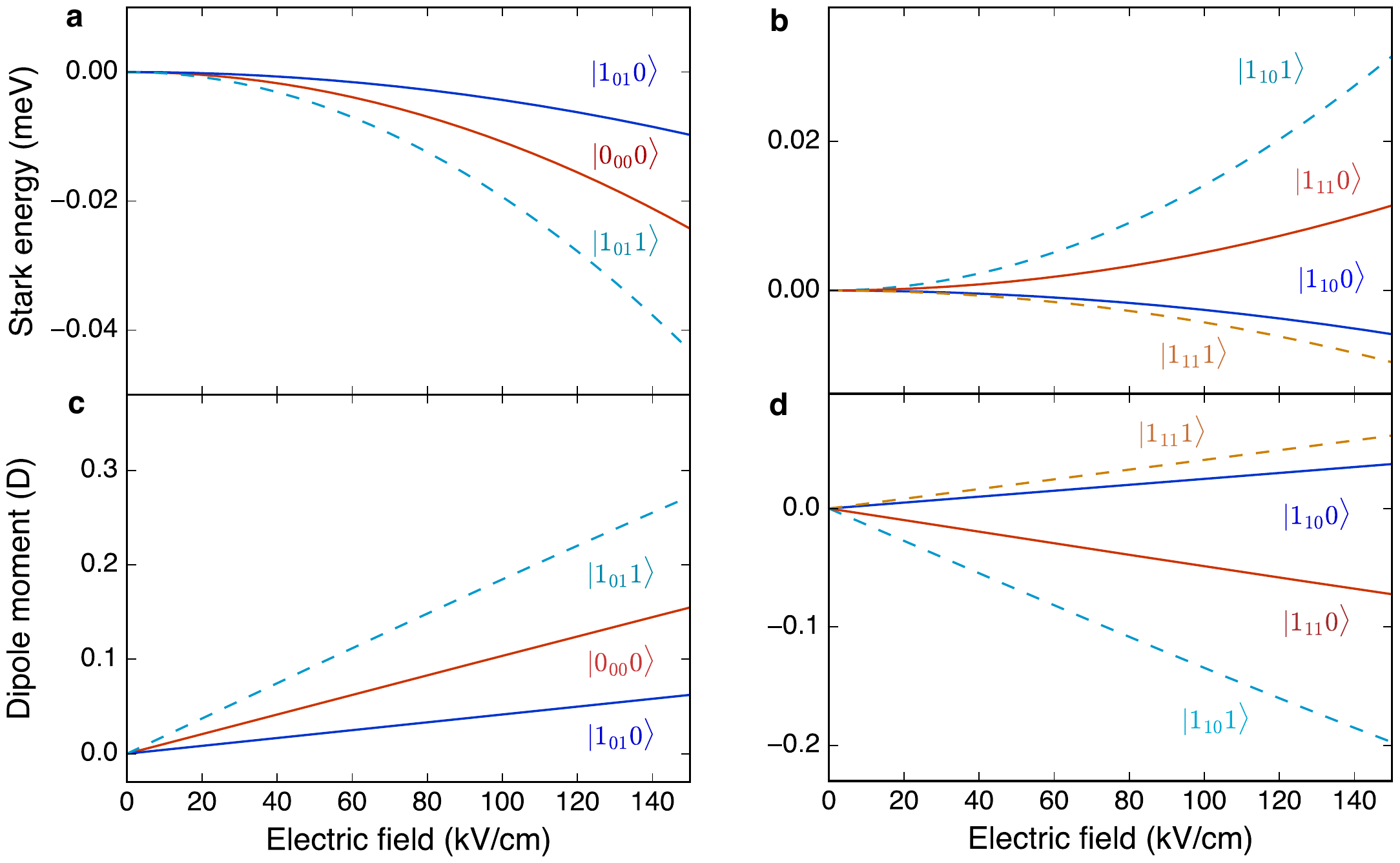}
\caption{Calculated stark energy shifts (a) and effective dipole moments (c) for the ground rotational states of \emph{ortho-} ($|j_{K_aK_c}|M|\rangle=|1_{01}0\rangle, |1_{01}1\rangle$) and \emph{para-} ($|0_{00}0\rangle$) water as a function of the electric-field strength. (b) and (d) Stark shifts and effective dipole moments for the higher excited rotational states $|1_{10}0\rangle, |1_{10}1\rangle$ and $|1_{11}0\rangle, |1_{11}1\rangle$.} 
\label{fig:stark}	
\end{figure}

Effective dipole moments and Stark energies of the individual rotational states of water were calculated using the CMIstark software package \cite{chang14a}. Figures \ref{fig:stark}a and c show simulated Stark energies and effective dipole moments, respectively, for the ground states of both isomers. All Stark components are high-field seeking with negative Stark shifts. The $|j_{K_aK_c}|M|\rangle=|1_{01}1\rangle$ state has the largest dipole moment resulting in the largest deflection in an electric field gradient, while $|1_{01}0\rangle$ shows the smallest Stark shift. Here, $M$ denotes the quantum number of the space-fixed projection of $\Vec{j}$. This results in a separation of the \emph{ortho-}water $|M|=0,1$ states under electrostatic deflection. The value of the dipole moment of \emph{para-}water in the $|0_{00}0\rangle$ state is in between those of \emph{ortho-}water. Additionally, calculations of the Stark energies and dipole moments for the excited rotational states $|1_{10}\rangle$ and $|1_{11}\rangle$ are presented in Figures \ref{fig:stark}b and d. Their contribution to the measured deflection profiles is discussed in Section \ref{sec:comp}.

The state-specific deflection profiles of the molecular beam traveling through the entire setup including deflector (Figure 1 of the main text) have been simulated using a home-made software package based on CMIfly \cite{chang15a}. The trajectory simulations were carried out with $10^6$ water molecules per quantum state. Their initial positions were uniformly sampled across the nozzle cross section. Their initial velocities were sampled from a normal distribution with a mean longitudinal velocity of 575 m/s and standard deviations of 10 m/s in the transverse directions and 69~m/s in the longitudinal direction. An \emph{ortho:para} ratio of 3:1 was imposed on the molecular beam set by the nuclear-spin statistics. For each quantum state, simulated deflection profiles were obtained from histograms of the coordinates of the molecules upon arrival in the reaction region. Thermally averaged deflection profiles $n_{\sigma, T}(y)$ ($\sigma \in \{o,p\}$) for each spin isomer at rotational temperature $T$ were calculated from the deflection profiles of the individual quantum states $n_{jK_aK_cM}(y)$ using the expression
\begin{equation}
n_{\sigma, T}(y) = \frac{g_\sigma}{N_\sigma} \sum_{j, K_a, K_c} \sum_{|M|=0}^{j} g_{M} g_{\sigma,K_aK_c} e^{-E_{jK_aK_c}/k_B T} \; n_{jK_aK_cM}(y) ,
\end{equation}
with the partition function
\begin{equation}
N_\sigma = \sum_{j, K_a, K_c} \sum_{|M|=0}^{j} g_{M} g_{\sigma, K_aK_c} e^{-E_{jK_aK_c}/k_B T}.
\end{equation}
Here, $k_B$ denotes the Boltzmann constant and $E_{jK_aK_c}$ are the field-free rotational energies. The degeneracy factor $g_M$ takes values $g_M = 1$ for $M=0$ and $g_M = 2$ for $|M|>0$ . The nuclear-spin symmetrization factors $g_{\sigma, K_aK_c}$ are defined by
\begin{equation}
g_{\mathrm{o}, K_aK_c} = \begin{cases} 1 & K_a + K_c \text{ odd} \\
0 & K_a + K_c \text{ even}\end{cases}
\end{equation}
for the \emph{ortho-} and 
\begin{equation}
g_{\mathrm{p}, K_aK_c} = \begin{cases} 0 & K_a + K_c \text{ odd} \\
1 & K_a + K_c \text{ even}\end{cases}
\end{equation}
for the \emph{para-}isomer.
The nuclear-spin degeneracy factors are $g_\text{p} = 1/4$ and $g_\text{o} = 3/4$. The total thermal deflection profile was calculated from the sum of the deflection profiles of the \emph{ortho-} and \emph{para-}isomers,
\begin{equation}
n_\mathrm{tot, T}(y) = n_{\mathrm{para}, T}(y) + n_{\mathrm{ortho}, T}(y).
\end{equation}

\section{Composition of the molecular beam}
\label{sec:comp}

\begin{figure}[h]
	\centering
		\includegraphics[width=0.6\textwidth]{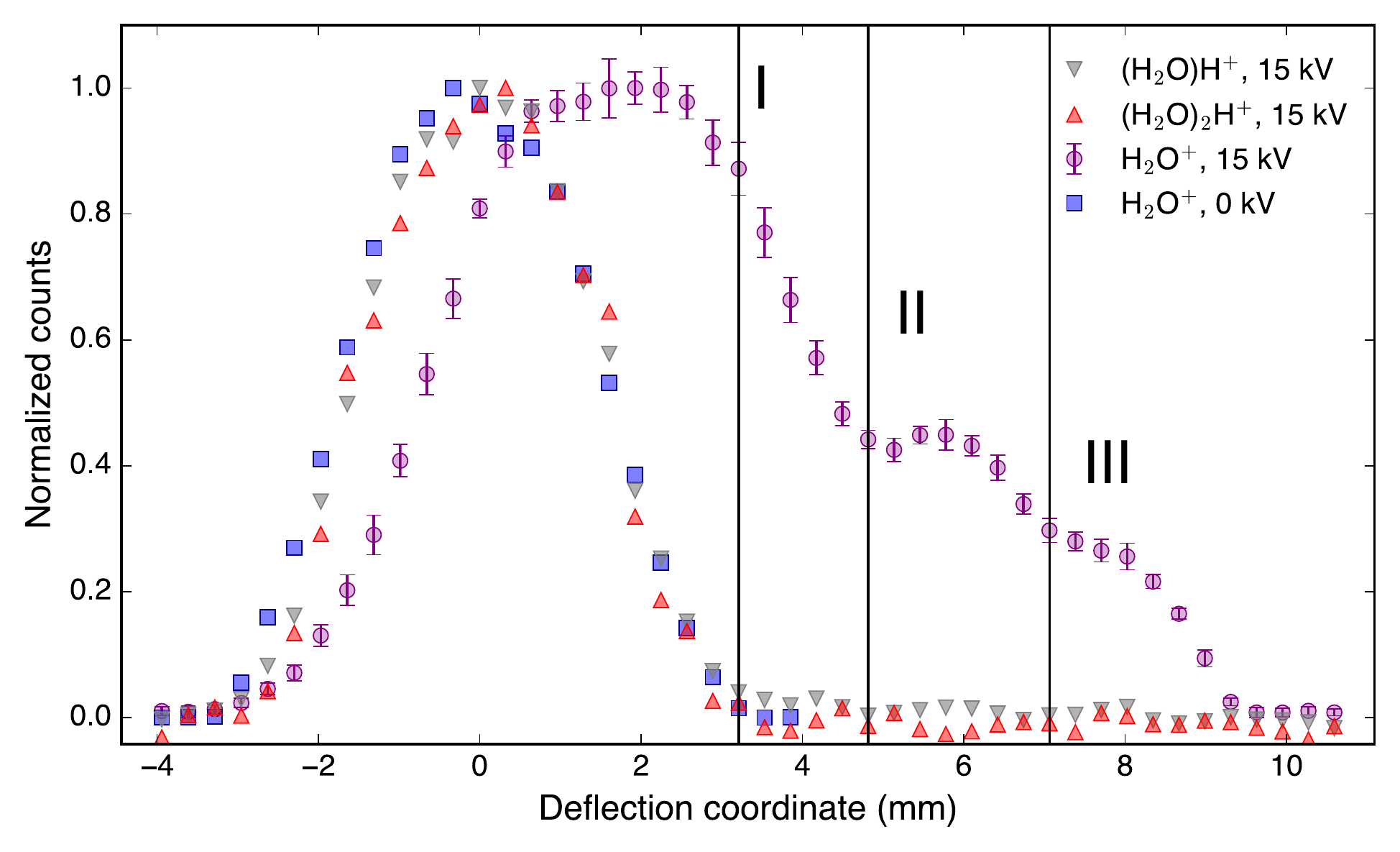}
 	\caption{Experimental deflection profiles of the water beam at a deflector voltage of 0 kV (blue squares) and 15 kV (purple circles) measured by fs-laser ionization. Deflection profiles of water clusters as detected by their fragment ions are also shown (grey and red triangles). These were produced from their parent molecules by fs-laser ionization and dissociation. No significant deflection is observed for these clusters. The vertical lines I, II and III specify the positions at which rate measurements were performed. Error bars correspond to one standard error. }
	\label{fig:SI_clusters}
\end{figure}

{\bf Contribution of water clusters:} The mass spectra of the molecular beam recorded after fs-laser ionization indicate the presence of water clusters in the supersonic expansion. The fs laser is able to ionize these clusters through strong-field ionization and break them into fragments including H$_2$O$^+$. The signal of these H$_2$O$^+$ cluster-fragment ions contributes to the $m/q=18$~u signal of the water monomer ions in the mass spectra. This additional signal needs to be taken into account when evaluating the measured deflection profiles of water. Figure \ref{fig:SI_clusters} shows that cluster fragments with mass $m/z>18$~u experience no significant deflection at a deflector voltage of 15 kV. We therefore assume that the detectable water clusters present in the molecular beam are largely undeflected and that the signal recorded at large deflection coordinates for $m/z=18$~u is solely due to the water monomer. Thus, to avoid interference from reactions of the N$_2$H$^+$ ions with the clusters, the measurement positions I, II and III specified in the main text were chosen outside the region of deflection coordinates where undeflected water clusters would contribute to the rate measurements. 

\begin{figure}[h]
	\centering
		\includegraphics[width=0.6\textwidth]{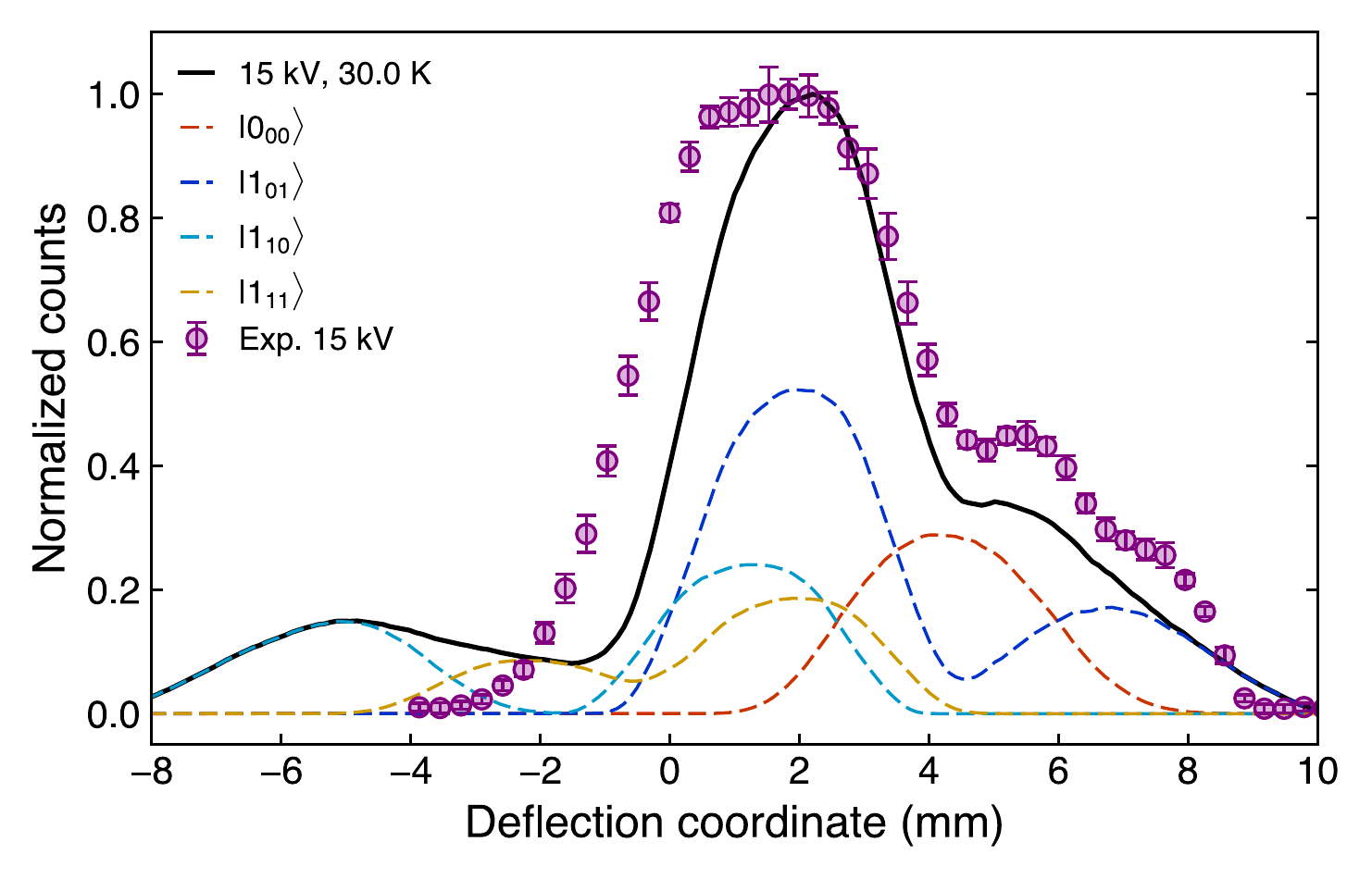}
 	\caption{Comparison of the deflection profile at 15 kV measured by fs laser ionization with a trajectory simulation at rotational temperature of 30 K. At this temperature, the partly low-field seeking states $1_{10}$ and $1_{11}$ are significantly populated. The simulation predicts a shoulder appearing at negative deflection coordinates. This is not observed in the experiment indicating that these higher rotational states are not significantly populated. Error bars correspond to one standard error. }
	\label{fig:SI_deflection_30K}
\end{figure}

\noindent {\bf Higher excited rotational states of water:} 
The REMPI spectrum in Figure \ref{fig:SI_rempi_scan_0kV} shows a weaker transition at 80664 cm$^{-1}$ which could indicate the population of the higher excited rotational state $|1_{10}\rangle$ in the molecular beam. To assess a possible contribution of higher rotational states to the deflection profiles, we show in Figure \ref{fig:SI_deflection_30K} simulations of a deflected beam at a rotational temperature of 30~K at which the $|1_{10}\rangle$ and $|1_{11}\rangle$ states are significantly populated. The Stark shifts and effective dipole moments of these states are shown in Figures \ref{fig:stark} b and d, respectively. In particular, these states exhibit low-field seeking Stark components which manifest themselves as shoulders in the deflection profile at negative deflection coordinates. As can be seen in Figures \ref{fig:SI_deflection_30K}, the experimental deflection profile exhibits no such features. We therefore conclude that higher excited rotational states do not play a significant role in the present experiments and that the measured reaction-rate constants predominantly reflect the contributions from the ground states of \emph{para-} and \emph{ortho-}water.


\section{Reaction-rate constants}
\label{sec:rate_constants}

{\bf Determination of pseudo-first order rate constants:} Pseudo-first-order reaction-rate constants were measured at the deflection coordinates I, II and III indicated in Figure 2b in the main text. At each of these positions, five independent sets of data were taken each consisting of one measurement with a deflected molecular beam (deflector voltage 15 kV) and one background measurement (deflector voltage set to 0 kV). For each data set, the N$_2$H$^+$ ion signal $N(t)$ was recorded as a function of reaction time $t$ in three consecutive measurements. The data were analyzed according to an integrated pseudo-first-order rate law
\begin{equation}
\ln\left(\frac{N(t)}{N(0)}\right) = - k' t
\end{equation}
with the rate constant $k'$ determined from a least-squares fit. For every data set, the pseudo-first-order rate constant of the reaction was calculated from the difference of the rate constants obtained with the molecular beam and the background measurement. 

\noindent {\bf Determination of the rate constants of $o/p$-H$_2$O:} From the measurements of the \emph{ortho}/\emph{para-} populations in the molecular beam using (2+1) REMPI and the pseudo-first-order rate constants at the three positions I, II and III, the pseudo-first-order rate constants $k'_o$ for \emph{ortho-} and $k'_p$ for \emph{para-}water were determined from the set of linear equations ($i \in \{ \rm I,II,III \}$) 
\begin{equation}
k'_{i} = \tilde{n}_{i} (p_{o,i}\;k'_o + p_{p,i}\;k'_p).
\label{eq:rates_equation_system}
\end{equation}
The relative beam densities are defined as $\tilde{n}_i = I_i/I_\mathrm{II}$ where $I_i$ is the normalized ion signal of the deflected molecular beam at deflection coordinate $y_i$ obtained by fs-laser ionization. The parameters defining the equations are summarized in Table \ref{tab:rates_equation_system}.
\begin{table}[h]
\centering
\begin{tabular}{l || c c c}
&  Pseudo 1$\rm^{st}$ oder rates & Populations & Normalized ion signal  \\
Region $i$ & $k'_i~[\mathrm{s}^{-1}]$ & $p_{o,i}$ : $p_{p,i}$ & $I_i$\\
\hline
I & $ 3.13(5)\cdot10^{-3}$ & $0.61(1)$ : $0.39(1)$ & $0.87(4)$\\ 
II & $ 1.63(4)\cdot10^{-3}$ & $0.38(2)$ : $0.62(2)$ & $0.44(1)$\\
III & $ 0.95(5)\cdot10^{-3}$ & $1$ : $0$ 			& $0.30(2)$\\
\end{tabular}
\caption{Total pseudo-first-order reaction-rate constants $k'_i$, nuclear-spin isomer populations $p_\sigma$ and normalzied ion signal $I_i$ at the three positions $i=$ I, II and III.}
\label{tab:rates_equation_system}
\end{table}

\noindent The pseudo-first-order rate constants for \emph{ortho-} and \emph{para-}water, $k'_o$ and $k'_p$, respectively, were obtained from a least-squares fit to the over-determined set of linear equations 
\eqref{eq:rates_equation_system} by minimising the cost function
\begin{equation}
S(k_o, k_p) = \sum_{i} (k'_i - \tilde{n}_{i} (p_{o,i}\;k_o + (1-p_{o,i})\; k_p))^2.
\label{eq:cost_function}
\end{equation}
The resulting pseudo-first-order reaction-rate constants are given in the first column of Table \ref{tab:rates_results}. 
The relative difference of the \emph{para-} and \emph{ortho-}rate constants was calculated using the expression
\begin{equation}
r = 2\frac{k_p - k_o}{k_p + k_o}.
\end{equation} 


\begin{table}[h]
	\centering
\begin{tabular}{l|| c c | c}			
& $k'$[s$^{-1}$] & $k$[cm$^3$s$^{-1}$] & $k_\mathrm{AC}$ [cm$^3$s$^{-1}$] \\ 
\hline
\emph{o-}H$_2$O & $1.4(1)\cdot10^{-3}$ & $4.8(9)\cdot10^{-9}$  & $4.0(6) \cdot 10^{-9}$ \\
\emph{p-}H$_2$O & $1.8(1)\cdot10^{-3}$ & $6(1)\cdot10^{-9}$ & $5(1)\cdot10^{-9}$ \\ 
Rel. difference $r$  &    \multicolumn{2}{c|}{$23(9)$\%} &  $24(5)$\% \\
\end{tabular}	
\caption{Experimental pseudo-first-order and second-order rate constants $k'$ and $k$, respectively, and theoretical adiabatic capture-rate constants $k_\text{AC}$ for the reaction of N$_2$H$^+$ with \textit{o-} and \textit{p-}H$_2$O. }

\label{tab:rates_results}
\end{table}

\section{Density of the molecular beam}

The time-averaged density of water molecules in the molecular beam, $n_\mathrm{MB,avg}$, relates to the density of molecules in one gas pulse, $n_\mathrm{MB,pulse}$, via
\begin{eqnarray}
n_\mathrm{MB,avg} &=& n_\mathrm{MB,pulse} f_\mathrm{rep} \tau_\mathrm{pulse},
\end{eqnarray}
where $f_\mathrm{rep}=200$ Hz is the repetition rate of the gas valve and $ \tau _\mathrm{pulse} = 487 \mu\mathrm{s}$  is the FWHM of the temporal width of the molecular beam pulse in the reaction region.

To estimate $n_\mathrm{MB,pulse}$, the fs-laser ionization signal $I_\text{II}$ of H$_2$O from the molecular beam obtained at deflection position II  was compared with the signal $I_\text{BG}$ obtained from ionization of water in the background gas of the UHV chamber. Assuming that the ionization volumes $V_\text{MB}$ and $V_\text{BG}$ were equal (the diameter of the molecular beam of 4~mm was larger than the laser-focus diameter of $\approx50~\mu$m), $n_\mathrm{MB,pulse}$ was estimated according to
\begin{equation}
n_\mathrm{MB,pulse} = n_\mathrm{BG}\frac{I_\mathrm{II}}{I_\mathrm{BG}} \underbrace{\frac{V_\mathrm{MB}}{V_\mathrm{BG}}}_{\approx 1} =  n_\mathrm{BG}\frac{I_\mathrm{II}}{I_\mathrm{BG}}
\end{equation}
Here, the density of water molecules in the background gas, $n_\mathrm{BG}$, was estimated from a comparison of the measured pseudo-first-order rate constant of the reaction of background water with N$_2$H$^+$ ions with a literature value for the second-order rate constant at 300~K:
\begin{equation}
n_\mathrm{BG} = k'_\mathrm{BG}/k_\mathrm{lit, 300 K},
\end{equation}
with $k_\mathrm{lit, 300 K} = 2.6(4)\times 10^{-9}~\mathrm{cm^{3}s^{-1}}$ taken from \cite{betowski75a}.
The value of the time-averaged beam density thus obtained is $n_\mathrm{MB, avg} = 3.0(5)\times10^5~\mathrm{cm}^{-3}$. This value is in line with previous results \cite{horke14a} after adjustment for the different repetition rates of the experiments.


\section{Collision velocity}
\label{sec:vcol}

In the present experiments, the collision energy was dominated by two contributions: the velocity of the H$_2$O molecules in the molecular beam and the velocities of the N$_2$H$^+$ ions in the Coulomb crystals.

\noindent {\bf Velocity of the molecular beam:} The velocity of the molecular beam was determined by probing its time-resolved density profile in the ion-trap region at different longitudinal positions of the gas valve. The velocity spread was determined from the half-width-at-half-maximum (HWHM) of the temporal profile of the gas pulse which was assumed to be limited by the longitudinal velocity distribution of the molecules. The velocity of the molecular beam was determined to be $v_{beam}=$575(50) m/s with a HWHM of $\delta v_{beam}=65(36)$~m/s. This beam velocity was used for the adiabatic-capture-rate calculations and the Monte Carlo trajectory simulations. 


\noindent {\bf Ion velocities:} The velocities of Coulomb-crystallized ions in RF traps are usually dominated by their micromotion, i.e., their fast motion driven by the RF fields \cite{willitsch12a}. The contribution of their secular motion can be neglected because of laser and sympathetic cooling. The spread of micromotion velocities across the crystal thus defines the velocity spread $\delta v$ of the entire ion ensemble. The micromotion velocities scale with the radial position of the ions in the trap \cite{major05a}. They are zero for ions located in the trap centre and maximal for the outermost ions of the crystal. The radial extension $r_0$ of N$_2$H$^+$ ions can directly be determined from the edges of the non-fluorescing central region of the crystal images taken during the measurements (see inset in Figure 1 of the main text). 
Using the analytical expressions of the ion trajectories in a quadrupole trap \cite{major05a}, the root-mean-square velocity for the outermost ions in the present N$_2$H$^+$ crystals at $r_0 = 200~\mu$m was estimated to be $v_\mathrm{rms}=163$ m/s. The corresponding micromotion energy was found to be $E_\mathrm{kin} =k_B\cdot47~$K. 
Thus, the total uncertainty in the experimental collision velocity was calculated to be
\begin{equation}
\delta v = \sqrt{v_{rms}^2+\delta v_{beam}^2} = 175~ \mathrm{m/s}.
\end{equation}

\section{Theory}

{\bf Adiabatic capture theory:} Rotationally adiabatic-capture rate constants were calculated using the theory developed by Clary and co-workers \cite{clary87a, stoecklin92a}. A collision velocity $v=575(175)$~m/s was assumed in the calculations. The long-range ion-molecule interaction potential adapted in the calculations consisted of the charge-induced dipole and charge-permanent dipole terms:
\begin{equation}
V(R, \beta) = - \frac{\alpha q^2}{2 R^4} - \frac{q \mu_D \cos\beta}{R^2},
\end{equation}
with $q$ the ion charge, $\alpha$ the molecular isotropic polarisability, $\mu_D$ the permanent dipole moment of the neutral molecule and $\mu$  the reduced mass of the collision partners. The ion-molecule distance is denoted $R$ and the orientation angle of the molecular dipole relative to the ion-molecule axis is denoted $\beta$. The molecular parameters for water used in this calculation are given in Table \ref{tab:molparams}.
\begin{table}[h]
\centering
\begin{tabular}{l| c l}			
    	Rotational constants 	&  $A$ & 27.88071 cm$^{-1}$\\
					& $B$ & 14.52181 cm$^{-1}$\\
					& $C$ & 9.27773  cm$^{-1}$\\\hline
	Reduced mass		& $\mu$ & 11.11533 u \\\hline
	Dipole moment		& $\mu_D$ & 1.85 D\\
	Polarisability		& $\alpha$ & $1.45 \text{\AA}^3$
\end{tabular}
\caption{Molecular parameters of water used in the adiabatic-capture calculations.}
\label{tab:molparams}
\end{table}

\noindent Following \cite{stoecklin92a}, a set of rotationally adiabatic, centrifugally corrected potential energy curves \linebreak $V_{JjK_aK_c\Omega}(R)$ was calculated (Figure 4b and c of the main text), where $J$ denotes the quantum number of total angular momentum of the system, $\Omega$ the quantum number of its projection on the intermolecular axis and $R$ the intermolecular distance. It is assumed that all collisions up to a maximum angular momentum $J_{max}$ for which the collision energy $E_\text{col}$ exceeds the centrifugal barrier of the effective potential lead to a successful reactive encounter. Hence, the reaction cross section is calculated to be \cite{stoecklin92a}
\begin{equation}
\sigma(j, K_a, K_c, E_c) = \frac{\pi \hbar^2}{2\mu E_c} \;\frac{1}{2 j + 1} \sum_{\Omega = -j}^{j} (J_\mathrm{max}(j, K_a, K_c, \Omega) + 1)^2.
\end{equation}
The state-selective reaction-rate constants are then given by $k(j, K_a, K_c, E_c) = \sigma(j, K_a, K_c, E_c) v$.

\noindent The calculated adiabatic-capture rate constants $k_\text{AC}$ for the ground states of the two nuclear-spin isomers of water and their relative difference $r$ are shown in table \ref{tab:rates_results} together with the experimental values. The quoted uncertainties of the theoretical rate constants reflect the spread $\delta v$ of the collision velocity $v$ (Section \ref{sec:vcol}) and were calculated from $\delta k = (k(v + \delta v) - k (v - \delta v))/2$.


\begin{figure}[h]
	\centering
		\includegraphics[width=0.6\textwidth]{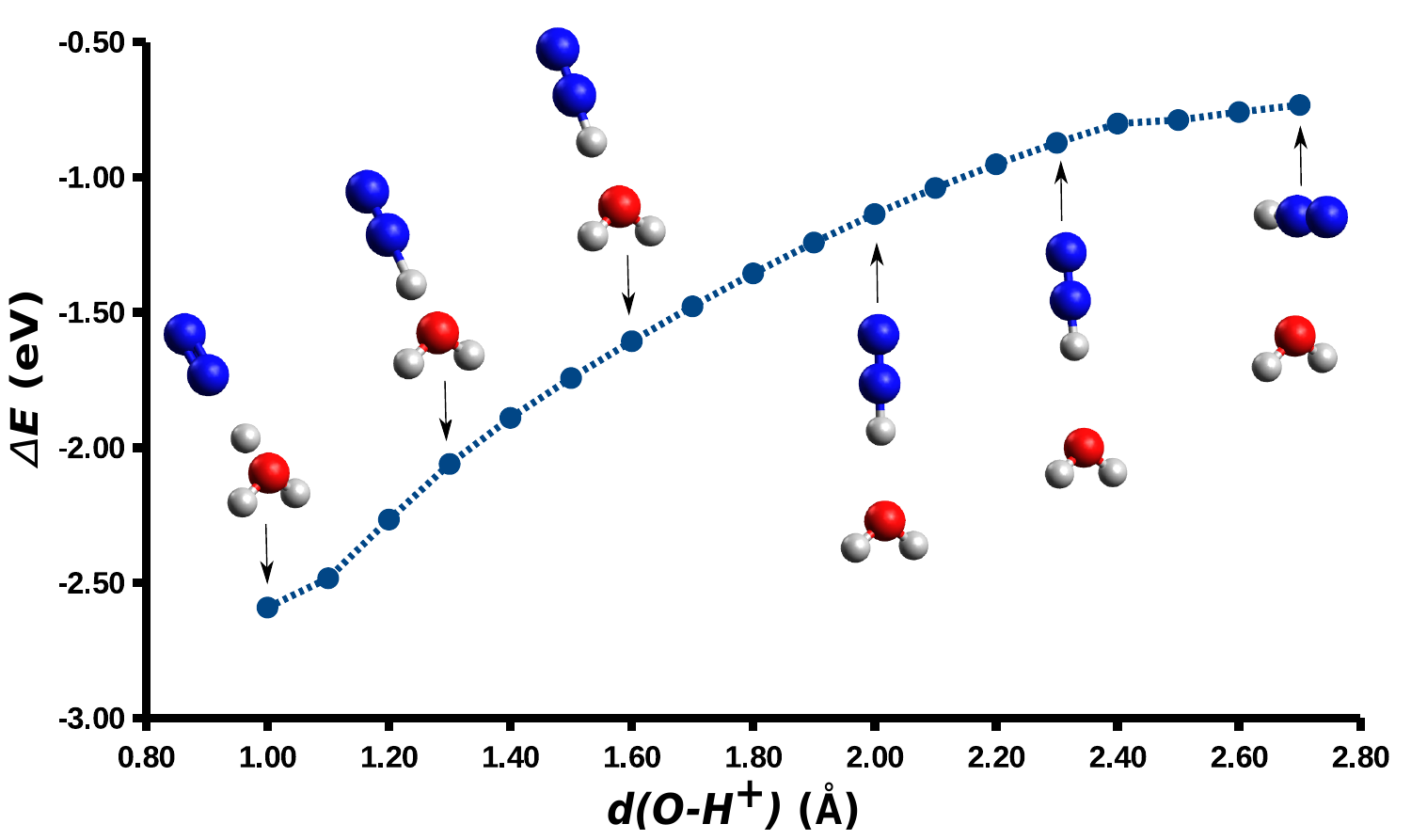}
 	\caption{Optimized structures along the reaction coordinate of the proton transfer between $\rm N_2H^+$ and $\rm H_2O$ at the CCSD/aug-cc-pVTZ level of theory. The energy is relative to that of the reactants. Blue spheres represent nitrogen atoms, red spheres oxygen atoms and white spheres hydrogen atoms.}
\label{fig:calc1}
\end{figure}
\noindent {\bf Quantum-chemical calculations:} In order to test whether the reaction $\rm H_2O + N_2H^+ \to H_3O^+ + N_2$ is indeed barrierless, electronic structure calculations were carried out. We assumed for this purpose that the O-H$^+$ distance approximates the reaction coordinate. We constrained this distance while optimizing all other internal coordinates at the CCSD/aug-cc-pVTZ level of theory. The resulting electronic energies for the optimized structures along the O-H$^+$ coordinate are shown in Figure \ref{fig:calc1}. Over the full range of O-H$^+$ distances investigated, the energy continuously decreases from the reactants towards the products (from right to left in Figure \ref{fig:calc1}). These results indicate that the reaction is indeed barrierless and vindicate the application of capture theory to model the reaction kinetics.

\bibliography{new_refs,Main-Oct17}
\bibliographystyle{apsrev}